\documentclass[journal, 10pt, letterpaper]{IEEEtran}

\usepackage[utf8]{inputenc}
\usepackage{amsmath}
\usepackage{amsfonts}

\usepackage{amsthm}

\usepackage{float}

\usepackage{graphicx}
\graphicspath{{Figures/}}
\usepackage{epstopdf}

\usepackage[font=scriptsize]{subcaption}
\DeclareCaptionFont{tiny}{\footnotesize} 
\usepackage{epsfig}
\usepackage{url}
\usepackage{comment}
\usepackage{graphics}
\usepackage{soul} 
\usepackage{multirow}

\usepackage{xargs}
\usepackage[colorinlistoftodos,prependcaption,textsize=small]{todonotes}
\newcommandx{\unsure}[2][1=]{\todo[linecolor=red,backgroundcolor=red!25,bordercolor=red,#1]{#2}}
\newcommandx{\change}[2][1=]{\todo[linecolor=blue,backgroundcolor=blue!25,bordercolor=blue,#1]{#2}}
\newcommandx{\info}[2][1=]{\todo[linecolor=OliveGreen,backgroundcolor=OliveGreen!25,bordercolor=OliveGreen,#1]{#2}}
\newcommandx{\improvement}[2][1=]{\todo[linecolor=Plum,backgroundcolor=Plum!25,bordercolor=Plum,#1]{#2}}
\newcommandx{\thiswillnotshow}[2][1=]{\todo[disable,#1]{#2}}

\usepackage{graphicx, blindtext} 

\usepackage{array}
\usepackage{ragged2e}
\newcolumntype{P}[1]{>{\RaggedRight\hspace{0pt}}p{#1}}

\usepackage{graphicx}

\usepackage{wrapfig}

\usepackage{algorithm}  
\usepackage{algorithmic}

\usepackage{listings}
\usepackage{psfrag,wrapfig}

\usepackage{enumerate}
\usepackage{subfiles} 

\newcommand{\eg}{{\it e.g.,}\xspace}
\newcommand{\viz}{{\it viz.,}\xspace}

\newcommand{\etal}{{\it et~al.}\xspace}
\newcommand{\ie}{{\it i.e.,}\xspace}
\newcommand{\etc}{{\it etc.}}
\newcommand{\ci}{{\it (i) }}
\newcommand{\cii}{{\it (ii) }}
\newcommand{\ciii}{{\it (iii) }}

\newcommand{\ca}{{\it (a) }}
\newcommand{\cb}{{\it (b) }}
\newcommand{\cc}{{\it (c) }}
\newcommand{\cd}{{\it (d) }}
\newcommand{\ce}{{\it (e) }}

\usepackage{multirow}

\usepackage{tikz}
\usetikzlibrary{arrows,shapes,positioning,shadows,trees}
\usepackage{forest}
\usetikzlibrary{arrows.meta}
\colorlet{linecol}{black!75}

\newcounter{myoptimizationproblemctr}

\soulregister\cite7

\begin{document}

\title{Securing Real-Time Internet-of-Things
	\thanks{\textsuperscript{*}These authors contributed equally to this work.}}

\author{\IEEEauthorblockN{Chien-Ying Chen\IEEEauthorrefmark{1}, Monowar Hasan\IEEEauthorrefmark{1} and Sibin Mohan}\\ \IEEEauthorblockA{Dept. of Computer Science, University of Illinois at Urbana-Champaign, Urbana, IL, USA}\\
	Email: \{cchen140, mhasan11, sibin\}@illinois.edu}

\maketitle





\documentclass[../rt_iot_main.tex]{subfiles}

	\begin{abstract}
Modern embedded and cyber-physical systems are ubiquitous. A large number of critical cyber-physical systems have real-time requirements (\textit{e.g.,} avionics, automobiles, power grids, manufacturing systems, industrial control systems, \textit{etc.}). Recent developments and new functionality requires real-time embedded devices to be connected to the Internet. This gives rise to the real-time Internet-of-things (RT-IoT) that promises a better user experience through stronger connectivity and efficient use of next-generation embedded devices. However RT-IoT are also increasingly becoming targets for cyber-attacks which is exacerbated by this increased connectivity. This paper gives an introduction to RT-IoT systems, an outlook of current approaches and possible research challenges towards secure RT-IoT frameworks.
	\end{abstract}


\section{Introduction}
Nowadays smart embedded devices 
(\eg surveillance cameras, home automation systems, smart TVs, in-vehicle infotainment systems, \etc) are connected to the Internet -- 
this rise in the Internet-of-things (IoT) links together devices/applications that were previously 
isolated.
On the other hand, embedded devices with \textit{real-time properties} (\eg strict \textit{timing} and \textit{safety} requirements) require interaction between cyber and physical worlds. These devices
are used to monitor and control physical
systems and processes in many domains, \eg manned and unmanned vehicles
including aircraft, spacecraft, unmanned aerial vehicles (UAVs), 
self-driving cars; critical infrastructures
; process control systems in industrial plants; 
smart technologies (\eg electric vehicles, medical devices, \etc)  
to name just a few. 
Given the drive towards remote monitoring and control, these devices are being increasingly interconnected, often via the Internet, giving rise to the {\em Real-Time Internet-of-things (RT-IoT)}. 
 Since many of these systems have to meet stringent \textit{safety} and \textit{timing} requirements,
any problems that deter from the normal operation of such systems could result in damage to the system, the environment or {\em pose a threat to human safety}. 
The drive towards remote monitoring and control facilitated by the growth of the Internet, 
the rise in the use of
commercial-off-the-shelf (COTS) components, standardized communication protocols
and the high value of these systems to adversaries are making cyber-security a design priority for such systems. Security breaches are not uncommon in critical IoT applications, especially considering the recent spate of IoT-centric attacks ( {\eg} the Marai botnet, attacks on the Dyn DNS provider, DoS attacks from IoT devices \cite{mirai, ddos_iot_camera}) as well as others centered on safety-critical systems ({\eg} Stuxnet \cite{stuxnet}, 
 BlackEnergy \cite{Ukraine16}, 
 attack demonstrations by researchers on
automobiles \cite{ris_rts_1, checkoway2011comprehensive} and medical devices
\cite{security_medical}.)
Successful cyber attacks against such
systems could lead to problems more serious than just loss of data or
availability because of their critical nature \cite{abrams2008malicious, checkoway2011comprehensive}. Attacks on one or more of these types of systems can have catastrophic results,
leading to loss of life or injury to humans, negative impacts on the system and even
the environment.

Enabling security in RT-IoT is often more challenging than generic IoT systems due to the additional real-time constraints.  The focus of this paper is to introduce the properties/constraints and security threats for RT-IoT (Sections \ref{sec:rt_iot_overview}-\ref{sec:threats}), summarize security solutions specially designed for such safety-critical domains (Section \ref{sec:current_approaches}) and highlight the research challenges (Section \ref{sec:discussion}). While there exit some surveys \cite{iot_sec_ind, ida2016survey, iot_sec1, iot_sec2, kraijak2015survey} on security and privacy issues in general-purpose IoT systems, to the best of our knowledge, there is no comprehensive summary in the context of RT-IoT security.





\section{Real-Time Internet-of-Things : An Overview} \label{sec:rt_iot_overview}

At their core, RT-IoT largely intersect with real-time
cyber-physical systems \cite{iot_control}. 
RT-IoT systems can be considered as a wide inner-connected network, in which nodes can be connected and controlled remotely. Table \ref{tab:rts} summarizes some
of the common properties/assumptions related to RT-IoT systems.
In this section, we intend to outline the elements of RT-IoT as well as the scope of security issues covered in this paper.
Figure~\ref{fig:rtiot-overview} gives some common scenarios where RT-IoT applications can be implemented.

\begin{figure}[ht]
\centering
\includegraphics[width=0.8\columnwidth]{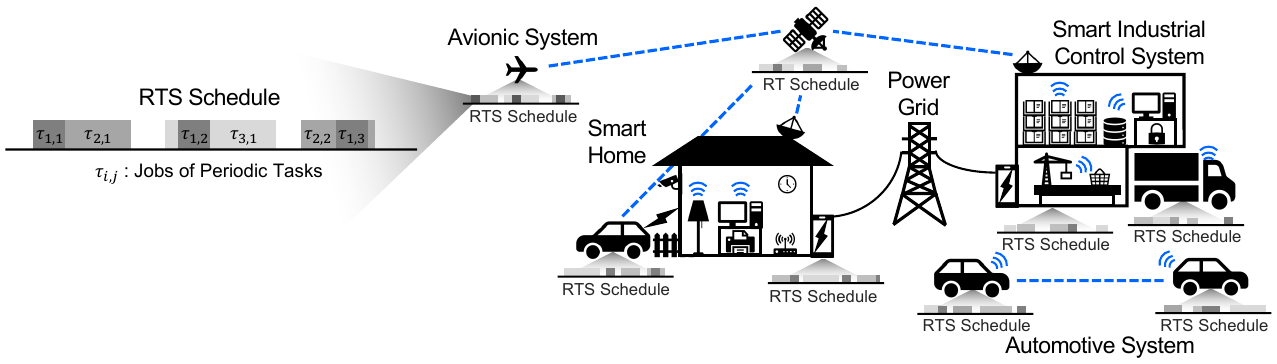}
\caption{An overview of RT-IoT around everyday living.
Dotted lines and radiate symbols indicate the wireless connectivity supported by the devices. Each RT-IoT device executes periodic real-time tasks (say $\tau_{i,j}$ -- that denotes the $j$-th activation of any task $\tau_i$) required for safe operation of the physical system.}
\label{fig:rtiot-overview}
\end{figure}

\subsection{Stringent Timing/Safety Requirements and Resources Constraints}




Many RT-IoT devices (\eg sensors, controllers, UAV, autonomous vehicles, \etc) will have severely limited resources (\eg memory,  processor, battery, \etc) and often require  control tasks to complete within a few milliseconds \cite{iot_fog_overview}. RT-IoT nodes, apart from a requirement for functional correctness,
require that temporal properties be met as well. These temporal properties are often
presented in the form of \textit{deadlines}. The usefulness of results produced by the system
drops on the passage of a deadline. If the usefulness drops sharply then we refer to
the system as a {\em hard} real-time system (\eg avionics, nuclear power plants, 
anti-lock braking systems in automobiles, \etc) and if it drops is a more gradual manner 
then they are referred to as {\em soft} real-time systems (\eg multimedia streaming, 
automated windshield wipers, \etc) \cite{Liu:2000:RTS}. 

\begin{table}[t h b] 
\centering
\caption{Properties of Majority RT-IoT Nodes}
\label{tab:rts}
\begin{tabular}{|c l|} \hline
$\bullet$ & Implemented as a system of periodic/sporadic tasks\\ \hline
$\bullet$ &  Stringent timing requirements\\ \hline
$\bullet$ & Worst-case bounds are known for all loops \\ \hline
$\bullet$ & No dynamically loaded or self modified codes \\ \hline
$\bullet$ & Recursion is either not used or statically bounded \\ \hline
$\bullet$ & Memory and processing power is often limited \\
\hline
$\bullet$ & Communication flows with mixed timing criticality \\
\hline\end 
{tabular}
\end{table}

\subsection{Heterogeneous Communication Traffic}
Many conventional RTS 
typically consist of several independently
operating nodes with limited or no
communication capabilities. However with the emergence of RT-IoT, cyber-physical nodes not only communicate over closed industrial
communication networks but are also often connected
via the Internet. 
Since most real-time applications would need to trigger events based on specific data conditions, a real-time communication
channel with guaranteed QoS (\eg throughput and data processing requirements, \textit{delay guarantees}, \etc) would also be necessary to support such applications \cite{rt-iot-ibm-china, rt-iot-qos-scheduling}.

Another property of RT-IoT is that they often
include traffic flows with mixed criticality, \ie those with varying degrees of
timing (and perhaps even bandwidth and availability) requirements: 
\ca {\em high priority/criticality traffic} that is essential for the
correct and safe operation of the system; examples could include sensors for closed loop control and
actual control commands in avionics, automotive or power grid systems; security systems in home automation 
\cb {\em medium criticality traffic} that is critical to the correct operation
of the system, but with some tolerances in delays, packet drops, \etc; for instance,
navigation systems in aircraft, system monitoring traffic in power substations, communication messages exchanged between electric vehicles and power grid or home charging station, traffic related to home automation equipment such as water sprinklers, heating, air conditioning, lighting devices, food preparation appliances \etc;
\cc {\em low priority traffic} -- essentially all other traffic in the system 
that does not really need guarantees on delays or bandwidth such as engineering traffic
in power substations, multimedia flows in aircraft,
notification messages from smart home equipment, \etc~
Typically, in many safety-critical RT-IoT, the properties of
all high-priority flows are well known, while the number and properties of other flows could
be more dynamic (\eg consider the on-demand video situation where new flows could
arise and old ones stop based on the viewing patterns of passengers in a commercial aircraft).




\subsection{Real-Time Scheduling Model}
Many such systems are implemented using a set of periodic (\eg fixed temporal septation between consecutive instances) or sporadic (\eg the tasks that can make an execution request at \textit{any} time, but with a \textit{minimum} inter-invocation interval) tasks \cite[Ch. 1]{rts_book}\cite{sporadic_task}. 
For instance, a sensor management task that monitors the conveyor belt in a manufacturing system needs to be periodic but the tasks that monitor the arrival of automated cars at traffic intersections are sporadic.
Another example is an engine control unit (ECU) in a modern vehicle in which the task that controls the valve in the electronic throttle body (ETB) is periodic while the task that handles commands from the in-vehicle computer is sporadic.
Application tasks in the RT-IoT nodes are often designed based on the Liu and Layland model \cite{Liu_n_Layland1973, rm_edf_review} that contains a set of tasks, $\Gamma$ where each task $\tau_i \in \Gamma$ has the parameters: $(C_i, T_i, D_i)$, where $C_i$ is the worst-case execution time (WCET), $T_i$ is the period or minimum inter-arrival time, and $D_i$ is the deadline, with $D_i \leq T_i$.


In the multicore context real-time task scheduling can be viewed as solving an allocation problem (\eg on which processor a task should execute) depending on design criteria \cite{multicore_survey}-- \eg~ \ci \textit{No migration:} tasks are allocated to a processor and no migration is permitted;
\cb \textit{Task-level migration:} the jobs of a task may execute on different core; however,
each job can only execute on a single core.
\cc \textit{Job-level migration:} The jobs of a task migrate to and execute on different cores;
however, parallel execution of a job is not permitted.

Schedulability tests \cite{res_time_1,res_time_rts,multicore_survey, bini2004schedulability} are used to determine if all tasks in the system meet their respective deadlines. If they do, then the task set is deemed to be \textit{`schedulable'} and the system, \textit{safe}.


\subsection{CPU Architectures and System Development Model}
\label{sec:architect}

Despite the fact that most RT-IoT applications are designed using platforms equipped with a single-core CPU, the trend towards multicore systems can be seen as many COTS devices nowadays are built on top of a multicore environment \cite{multicore_survey}.
For some specific applications (\eg avionics systems), there exist regulations that restrict the use of additional cores.
In such cases, the additional cores that do not execute real-time or safety critical tasks can be utilized to provide layers of security to the system. We have leveraged the use of multicore platforms in the real-time domain and developed security solutions \cite{mohan_s3a,securecore,securecore_memory,securecore_syscal,onchip_fardin,mhasan_resecure16} as discussed in Section \ref{subsec:sec_w_hw}.


It is also common that multiple vendors are involved in the development of RT-IoT systems.
Such a system is said to be developed under the \textit{multi-vendor development model}~\cite{sg2}.
In this model, each vendor designs/controls several separate tasks.
Figure~\ref{fig:vendor-based-model} demonstrates an electronic control unit (ECU) for an avionics system (on an unmanned aerial vehicle) that uses the multi-vendor development model. While this demonstrative example focuses on the avionics domain other RT-IoT systems (\eg automotive, home automation, \etc) could also be created using a similar model (albeit loosely defined).
\begin{figure}[ht]
\centering
\includegraphics[width=0.8\columnwidth]{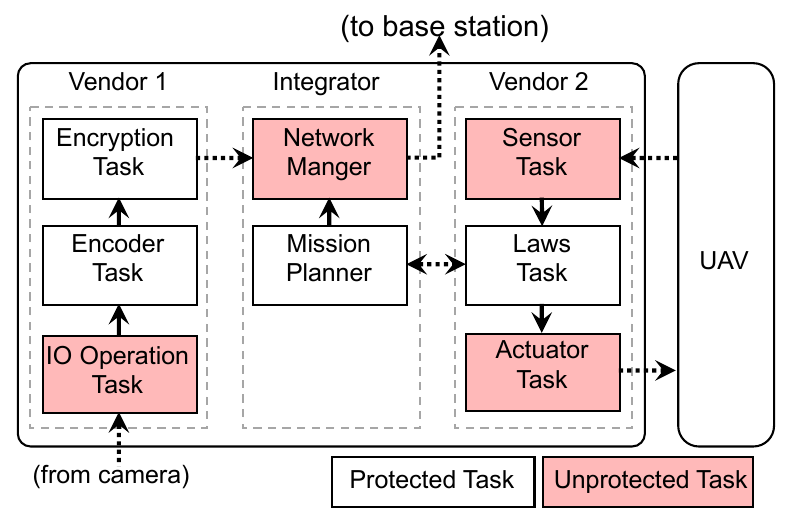}
\caption{A high-level design of a UAV that exemplifies the multi-vendor development model. In this demonstrative system, three vendors are involved in building the ECU system --
\textsl{Vendor 1} comprises tasks that process image data from a surveillance camera attached to the ECU;
\textsl{Vendor 2} is in charge of flight control tasks interacting with the UAV;
\textsl{Integrator} handles communication between the system and a base station.}
\label{fig:vendor-based-model}
\end{figure}



\section{Security Threats for RT-IoT}
\label{sec:threats}

RT-IoT systems face threats in various forms depending on the system and the goals of an adversary. In a system developed using vendor-based model, one of the involved vendors can act maliciously.
This (potentially unverified/untrusted) vendor could embed malicious functions in its tasks.
Bad coding practices could also leave vulnerabilities even if the involved vendors are {\em not} malicious. leveraging such system vulnerabilities adversaries can execute malicious codes (Section \ref{subsec:codeinject}), infer critical system information (Section \ref{subsec:sidechannel}) and/or perform denial of service attacks (Section \ref{subsec:dos}).
In a system that has network connectivity, the adversary could target the communication interfaces (Section \ref{subsec:comm}).
Due to a lack of authentication in many of these systems, the communication channels could easily be intercepted and forged.

\subsection{Attacks on RT-IoT}
We classify the attack methodologies on RTS based on the control over computational processes and the functional objective of the attack. One way to acquire control over a target system could be the injection of malicious code (\eg malware) or by reusing legitimate code for malicious purposes (\eg \textit{code-injection attacks}). Besides, since RT-IoT nodes can communicate over unreliable mediums such as Internet, the system is also vulnerable to \textit{network-level attacks}. 
Other than trying to aggressively crash the system (\eg using \textit{DoS attacks}) the adversary may silently lodge itself in the system and extract sensitive information (\eg \textit{side-channel attacks}). 
The side-channel attacks are based on observing properties (\eg execution time, memory usage patterns, task schedule, power consumption, \etc) of the system. 
This information may later be used by the attacker to launch further attacks. 
In the rest of this section, we summarize the common attack surfaces for RT-IoT systems.


\subsubsection{Integrity Violation with Malicious Code Injection} \label{subsec:codeinject}
An intelligent adversary can get a foothold in the system. For example, an adversary may insert a malicious task that respects the real-time guarantees of the system to avoid immediate detection and/or compromise one or more existing real-time tasks. The attacker may use such a task to manipulate sensor inputs and actuator commands (for instance) and/or modify system behavior in undesirable ways. Integrity violation through code injection attacks conceptually consists of two steps \cite{code-injection-sensor}. First, the attacker sends instruction snippets (\eg a valid machine code program) to the device that is then stored somewhere in memory by the software application receiving it. Such instruction snippets are referred to as 
\textit{gadgets}. In the second step, the attacker triggers a vulnerability in the application software, \ie real-time OS (RTOS) or task codes to divert the control flow. Since the instruction snippets represents a valid machine code program, when the program execution jumps to the start address of the data, the malicious code is executed. As we illustrate in Section \ref{sec:current_approaches} our recent solutions \cite{mohan_s3a,securecore,securecore_memory,securecore_syscal,onchip_fardin,mhasan_resecure16,mhasan_rtss16,mhasan_ecrts17,mhasan_date18} can be used to detect integrity violations through a combination of hardware/software mechanisms.

\subsubsection{Side-Channel Attacks}
\label{subsec:sidechannel}
The adversary may learn important information by side or covert-channel attacks \cite{sidechannel_survey} by simply lodging themselves in the system and extracting sensitive information. 
A side-channel attack manipulates previously unknown channels to acquire useful information from the victim.
Memory/cache access time~\cite{kelsey1998side}, power consumption traces~\cite{Jiang2014}, schedule preemptions~\cite{embeddedsecurity:son2006}, electromagnetic (EM) emanations~\cite{agrawal2002side} and temperature~\cite{bar2006sorcerer} \etc~are examples of some typical side-channels used by attackers.
These attack surfaces are particularly applicable to attacking RT-IoT nodes that execute real-time tasks due to the deterministic behaviors in such systems. 
A demonstrative cache-timing attack is presented in Section~\ref{sec:cache-attack-demo} and
Section~\ref{sec:secure-side-channel} illustrates our recent approaches \cite{sg1,sg2} to mitigate information leakage that used timing-based attacks on storage-channels.

\subsubsection{Attacks on Communication Channels}\label{subsec:comm}

RT-IoT elevates the Internet as the main communication medium between the physical entities. However,  Internet, as a insecure communication medium, introduces a variety of vulnerabilities that may put the security and privacy of RT-IoT systems under risk. Threats to communication includes eavesdropping or interception, man-in-the-middle attacks, falsifying, tampering or repudiation of control/information messages~\cite{loukas2015cyber}.
From the perspective of RT-IoT,  defending against communication threats is not an easy task. This is because it is challenging to distinguish rogue traffic from the legitimate traffic (especially for the critical/high-priority flows) without degrading the QoS (\eg bandwidth and end-to-end delay constraints). Threats to communications are usually dealt by integrating cryptographic protection mechanisms \cite{xie2007improving,lin2009static}. However this increases the WCET of the real-time tasks and may require modification of existing schedulers. Many cryptographic operations are also computationally expensive to execute especially on limited resources available in embedded RT-IoT devices. 
Therefore existing cryptographic approaches may not be a preferable option for many RT-IoT systems. In Section \ref{subsec:sec_int} we illustrate a solution to integrate security mechanisms that can also be used for dealing with communication threats but does not require modification of existing real-time tasks.

\subsubsection{Denial-of-Service (DoS) Attacks}\label{subsec:dos}

Due to resource constraints (\eg low memory capabilities, limited computation resources, \etc) and stringent timing requirements, RT-IoT nodes are vulnerable to DoS attacks. The attacker may take control of the real-time task(s) and perform system-level resource (\eg CPU, disk, memory, \etc) exhaustion. A more severe type of the DoS attack is the distributed denial-of-service (DDoS) attack where a large number of malicious/compromised nodes simultaneously attack the physical plant. 
In particular, when critical tasks are scheduled to run, an attacker may capture I/O or network ports and perform network-level attacks to tamper with the confidentiality and integrity (\viz safety) of the system. Again the defense mechanisms developed for generic IT or embedded systems do not consider timing, safety and resource constraints of RT-IoT and are not easily adaptable without significant modifications. As described in Section \ref{subsubsec:resecure} and \ref{subsec:sec_int}, our recent work \cite{mhasan_resecure16, mhasan_rtss16,mhasan_ecrts17,mhasan_date18} may be used to defend against DoS attacks.


But first, in order for those attacks to be successful, \textit{reconnaissance} is one of the early steps that an attacker needs to carry out. We illustrate this in the following (to demonstrate an attack mechanism).

\subsection{Reconnaissance: Attack Preparation}
\label{sec:reconnaissance}

Reconnaissance, essentially, is the first step for launching other successful attacks and, at the very least, the attacker gains important information about the system's internals.

\subsubsection{ScheduLeak}
\label{sec:scheduleak}
In initial work \cite{chen2015schedule}, we developed an algorithm, 
``\textit{ScheduLeak}'', to show the feasibility of a {\em schedule-based side-channel attack targeting real-time embedded systems} with a multi-vendor development model introduced in Section~\ref{sec:architect}.
The adversary 
could be one of the vendors or an attacker who compromises a vendor.
The \textit{ScheduLeak} algorithm utilizes an {\em observer task} that has the lowest priority in the victim system to observe \textit{busy intervals}.
A ``busy interval'' is a block of time when one or more tasks are executing -- an adversary \textit{cannot} determine {\em what} tasks are running {\em when} by just measuring or observing the busy intervals as they are. 

The \textit{ScheduLeak} algorithm can be represented as a function
$R(\Gamma, W)={J}$,
where $W$ is a set of observed busy intervals and $J$ is the inferred schedule information that can be used to pinpoint the possible start time of any particular victim task. 
Such a function is illustrated by Fig.~\ref{fig:scheduleak}.
%
By using the \textit{ScheduLeak} algorithm, an attacker can deconstruct the observed busy periods 
(with up to 99\% success rate if tasks have fixed execution times) 
into their constituent jobs and precisely pinpoint the instant when a task is scheduled.

\begin{figure*}[t h]
    \centering
    \begin{subfigure}[t]{1\textwidth}
        \centering
     \includegraphics[width=0.55\columnwidth]{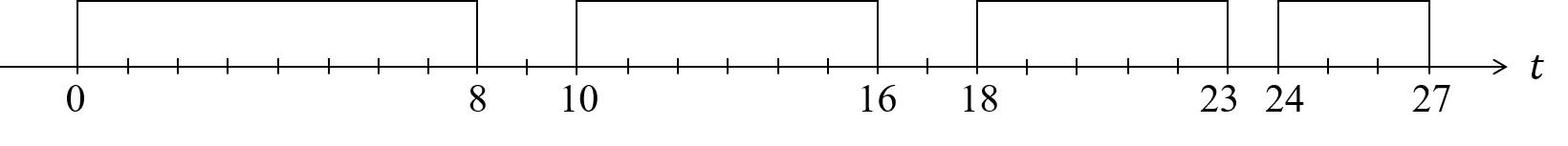}
\vspace{-0.5\baselineskip}
        \caption{Busy intervals observed by attacker's observer task.}
    \end{subfigure}%
    \hfill  
    \vspace{0.5\baselineskip}
    \centering
    \begin{subfigure}[t]{1\textwidth}
        \centering
        \includegraphics[width=0.55\columnwidth]{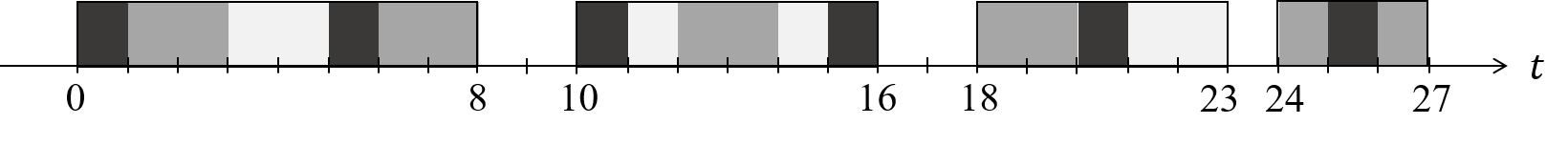}
        \caption{Schedules reconstructed by the \textit{ScheduLeak} algorithm.}
    \end{subfigure}
    \caption{An example of the schedules produced from a task set of three tasks \cite{chen2015schedule}. The \textit{ScheduLeak} algorithm can recover the precise schedules from the observed busy intervals.}
    \label{fig:scheduleak}
\end{figure*}

\subsubsection{Targeted Attacks}
\label{sec:cache-attack-demo}
It's worth mentioning that the effectiveness of side-channel attacks is enhanced when combined with the reconnaissance step we just introduced.
For example, in the demonstrative ECU system introduced in Section~\ref{sec:architect}, let us assume code inserted into \textsl{Vendor 2} would like to identify whether the surveillance camera controlled by the I/O Operation Task is enabled.
The attacker can launch a \textit{ScheduLeak} algorithm to infer exact start times of the IO Operation Task and carry out a cache-timing attack to gauge cache usage when an I/O Operation Task is scheduled.
Figure~\ref{fig:cache_attack} shows the result of such a cache-timing attack.
By launching a \textit{ScheduLeak} attack and knowing when the I/O Operation Task is scheduled to execute, the attacker probes the cache usage only when the task is active.
The result indicates that the attacker is able to identify the instant when the camera is on (\ie when a large amount of data is processed by I/O Operation Task).

\begin{figure}[h]
\centering
\includegraphics[width=0.68\columnwidth]{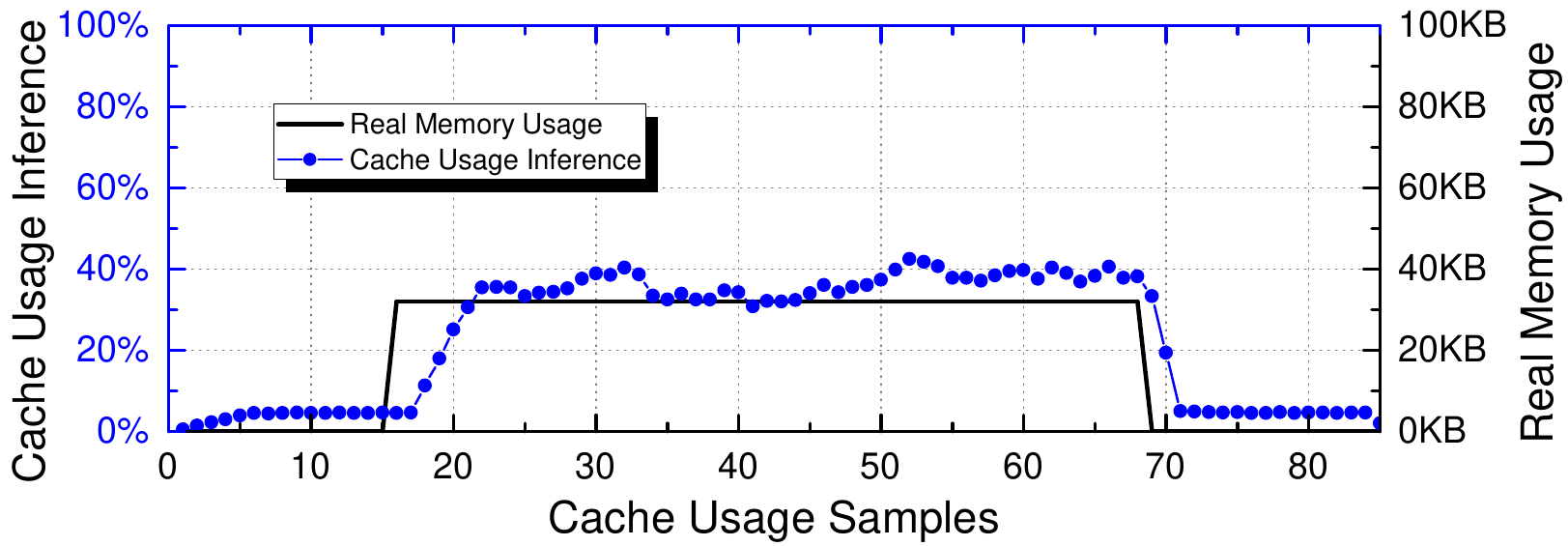}
\caption{A demonstration of a cache-timing attack \cite{chen2015schedule}. The X-axis is sample points and Y-axis shows both cache usage inference (round dots) and real memory usage amount (the solid line). It shows that a successful cache-timing attack can precisely infer the memory usage of the victim task.}
\label{fig:cache_attack}
\end{figure}




\section{Securing RT-IoT: Host-based Approaches}
\label{sec:current_approaches}

In what follows we summarize our initial attempts to provide security in RT-IoT nodes. We refer to these approaches as \textit{host-based} solutions since they primarily focus on securing an individual RT-IoT node. 
These approaches can be classified into two major classes: \ci solutions that require custom hardware support to provide security and \cii the solutions at the scheduler/software level that do not require any architectural modifications. Table \ref{tab:attack_n_sol} summarizes these security mechanisms for RT-IoT systems.

\begin{table*}[!htb]
\renewcommand{\arraystretch}{1.3}
\caption{Summary of Security Solutions for RT-IoT}
\label{tab:attack_n_sol}
\centering
\begin{tabular}{P{2.3cm}|P{5cm}|P{3.2cm}|P{3.7cm}}
\hline
\bfseries Reference & \bfseries Approach & \bfseries Attack Surface & \bfseries Overhead/Costs\\
\hline\hline
Simplex-based security \cite{mohan_s3a, securecore, securecore_memory, securecore_syscal, onchip_fardin} & Use verified/secure hardware module to monitor system behavior (\eg timing \cite{securecore} and execution pattern \cite{mohan_s3a}, memory access \cite{securecore_memory}, system call usage \cite{securecore_syscal}, control flow \cite{onchip_fardin}) & Code injection attacks & Require custom hardware or monitoring unit\\ \hline
Security by platform-level reset \cite{mhasan_resecure16, abdi2018guaranteed} & Periodically and/or asynchronously (\eg upon detection of a malicious activity) restart the platform and load an uncompromised OS image & Code injection, side channel and DoS attacks & Extra hardware to ensure safety during periodic/asynchronous restart events\\ \hline
Cache flushing \cite{sg1,sg2} & Flush the shared medium (\eg cache) between the consecutive execution of high-priority (security sensitive) and low-priority (potentially vulnerable) tasks & Side-channel (cache) attacks & Overhead of cache flushing reduces task-set schedulability \\ \hline
Schedule randomization \cite{taskshuffler}& Randomize the task execution order (\ie schedule) to reduce the predictability & Side-channel attacks & Extra context switch \\ \hline
Security task integration for legacy RT-IoT \cite{mhasan_rtss16,mhasan_date18} & Execute monitoring/intrusion detection tasks with a priority lower than real-time task to preserve the real-time task parameters (\eg period, WCET and execution order)  & Code injection, side-channel, DoS and/or communication attacks depending on the what monitoring tasks are used & Running security task with lower priority may cause longer detection time due to high interference (\eg preemption) from real-time tasks \\
\hline
Adaptive security task integration \cite{mhasan_ecrts17} & Execute monitoring/intrusion detection tasks with a lowest priority most of the time (\eg during normal system operation) -- however change the mode of operation execute with a higher priority (for a limited amount of time) if any anomalous behavior is suspected  & Code injection, side-channel, DoS and/or communication attacks depending on the what monitoring tasks are used & False positive detection may cause unnecessary mode switches \\
\hline
\end{tabular}
\end{table*}

\subsection{Security with Hardware Support}
\label{subsec:sec_w_hw}

The key idea of providing security \textit{without} compromising the \textit{safety} of the physical system is built on the
\textit{Simplex} framework \cite{sha2001using}. Simplex is a well-known real-time architecture that utilizes a minimal, verified controller (\eg \textit{safety controller}) as backup when the complex, high-performance controller (\eg \textit{complex controller}) is not available or malfunctioning. The goal of the Simplex method is to guarantee that even though a safety-critical system is controlled by a complex controller, the physical system would remain \textit{safe}. We have used the idea of Simplex  in the context of RT-IoT security \cite{mohan_s3a, securecore,securecore_memory,securecore_syscal,mhasan_resecure16}. The key concept of using Simplex-based architecture for security is to use a minimal simple subsystem (say a trusted core) to \textit{monitor} the properties (\ie timing behavior \cite{mohan_s3a, securecore}, memory access \cite{securecore_memory}, system call trace \cite{securecore_syscal}, behavioral anomalies \cite{mhasan_resecure16}, \etc) of an untrusted entity (\eg monitored core) that is designed for more complex tasks and/or exposed to less secure mediums (\eg network, Internet, I/O channels, \etc).

\subsubsection{Secure System Simplex Architecture (S3A)}

As mentioned in Section \ref{sec:rt_iot_overview}, the worst-case, best-case and average-case behaviors
for most RT-IoT nodes are calculated ahead of time to ensure
that all resource and schedulability requirements will be met
during system operation. 
\textit{S3A} \cite{mohan_s3a} utilizes this knowledge of deterministic execution profile of the system
and use to detect the violation of predicted (\eg uncompromised) system behavior.  \textit{S3A} is one of our earliest efforts to use another (FPGA-based, in this case) trusted
hardware component that monitors the behavior (\eg \textit{execution time} and the \textit{period}) of
a real-time control application running on a untrustworthy main
system.  The goal of this Simplex-based architecture is to detect an infection as quickly as possible and then
ensure that the physical system components always remain safe. Using an FPGA-based implementation and considering inverted pendulum (IP) as the physical plant we demonstrated that \textit{S3A} can detect intrusions in less than 6 $\mu s$ 
without violating safety requirements of the actual plant.

\subsubsection{SecureCore Framework}

\begin{figure}
\centering
\includegraphics[width=0.85\columnwidth]{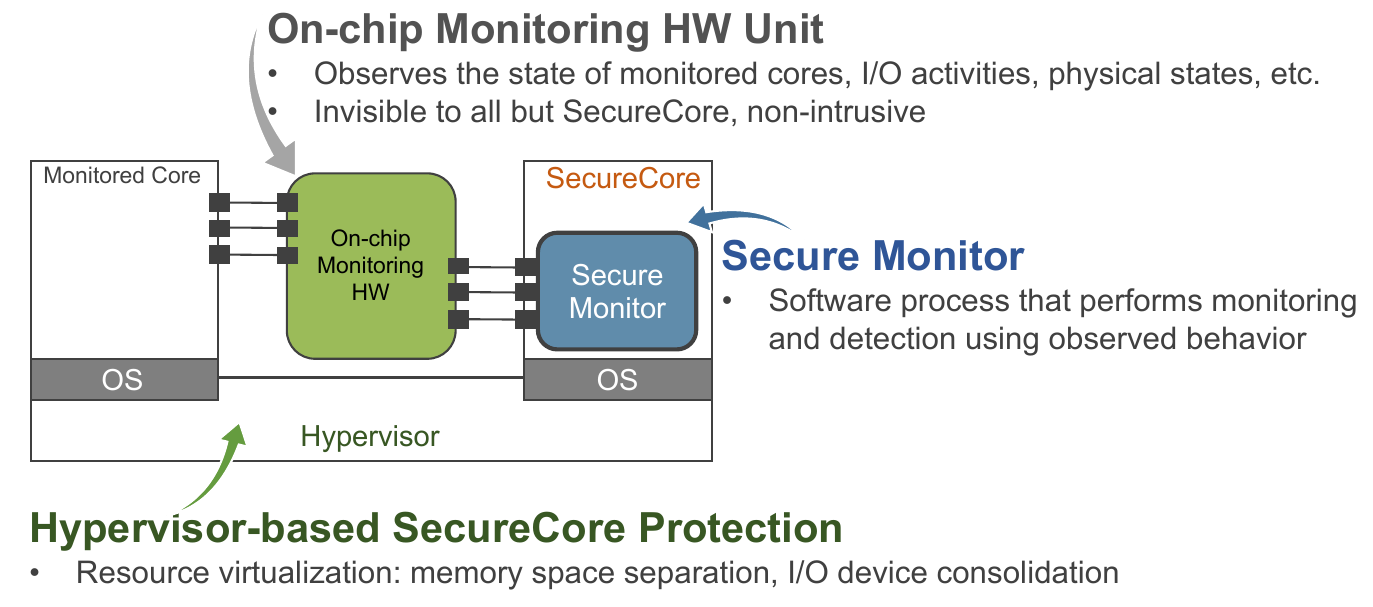}
\caption{An illustration of \textit{SecureCore} framework. The trusted core is used to monitor the behavior of the complex (and potentially vulnerable) core used for executing application/control tasks.}
\label{fig:securecore}
\end{figure}

As illustrated in Fig. \ref{fig:securecore} the idea of \textit{SecureCore} architecture is to utilize the redundancy in multicore chips to create a trusted entity (\eg a
`secure' core) that can continuously monitor the system behavior (\eg code execution pattern \cite{securecore}, memory usage \cite{securecore_memory}, system call trace \cite{securecore_syscal})
of a real-time application on an untrustworthy entity (\eg monitored core). The SecureCore is protected by hypervisor-based approaches (\eg by isolating memory regions and I/O device consolidation). The secure monitor (a software process) in the SecureCore uses the on-chip hardware monitoring unit to observe the states (\eg I/O activities, memory usages, \etc) of monitored cores and checks the system behavior at runtime.

The initial \textit{SecureCore} architecture \cite{securecore} uses a statistical learning-based mechanism for profiling the correct
execution behavior of the target system and uses these profiles to detect
malicious code execution. Given the probability distribution $P(e)$ of a legitimate
execution instance, the secure monitor compares $P(e)$ with a predefined minimum
required probability $\theta$  --- if
$P(e)$ is below the threshold probability (\eg $P(e) < \theta$) the execution instance to
is considered as malicious. 
The \textit{SecureCore} framework is also extended \cite{securecore_memory} to profile memory behavior (referred to as memory heat
map (MHM)) and then detect deviations from the normal memory
behavior patterns. MHM represents
how many times a particular memory region was accessed
during a
time interval. We proposed machine learning algorithms to 
characterize the information contained in the
MHMs and then detect deviations from the normal memory
behavior patterns. We have also extended \textit{SecureCore} architecture to detect anomalous executions using a distribution of system call
frequencies. Specifically we have proposed \cite{securecore_syscal} to use clustering algorithms (\eg global $k$-means clustering
with the Mahalanobis distance) to learn the legitimate execution
contexts (by means of distribution
of system call frequencies) of real-time applications and then monitor
them at run-time to detect intrusions.

\subsubsection{Control Flow Monitoring}

We then proposed hardware-based approach for checking the integrity of code flow of real-time tasks \cite{onchip_fardin}. In particular, we 
add an on-chip control flow monitoring module (OCFMM)
with a dedicated memory unit that directly hooks  into the processor and tracks the control
flow of the tasks.  The control flow graph (CFG) of tasks is produced
from the program binary and loaded into the OCFMM memory in advance (\eg during system boot). 
The detection module inside OCFMM
compares the control flow of the running program with the stored one (\eg CFG profiles that are
loaded into the dedicated memory at boot time) during program execution. At run-time (\eg during execution of a given block) CFG profiles
for the next-possible blocks are pre-fetched. The decision module continuously
scans the  current
block and validates the execution flow by comparing the current address
of the program counter (PC) against the possible, previously fetched
destination addresses. If any mismatch
occurs, the detection module raises a \textit{detection flag} that indicates a possible breach.

\subsubsection{Security via Platform-level Reset}
\label{subsubsec:resecure}

In traditional computing systems (\eg servers, smart phones, \etc), software problems are often resolved by restarting either the application process or the platform~\cite{candea2003jagr, candea2001recursive}. However, unlike those conventional computing systems restart-based recovery mechanisms are not straightforward in RT-IoT due to the real-time constraints as well as interactions of the control system with the physical world (for example, a UAV can quickly be destabilized if its controller restarts). In initial work \cite{mhasan_resecure16} we proposed a restart-based concept to improve security guarantees for RT-IoT. This Simplex-based framework, that we refer to as \textit{ReSecure}, is specifically designed to improve security of safety-critical RT-IoT systems. In particular, we propose to \textit{restart} the platform periodically/asynchronously and load a fresh image of the applications and OS from a \textit{read-only media} after each reboot with the objective of \textit{wiping out} the intruder or malicious entity. 
The ReSecure architecture (see Fig.~\ref{fig:resecure}) produces a verified system (by using a safety unit) despite the use of an unverified complex controller (\eg complex unit). 
OS/Firmwire in 
complex unit is exposed to external (possible attack) surfaces and can fail. Decision module predicts if the future states are safe. Watchdog (WD) and periodic timers restart the complex unit (and reload OS image from read-only memory) upon fail-stop.

\begin{figure}[t]
\centering
\includegraphics[scale=0.38]{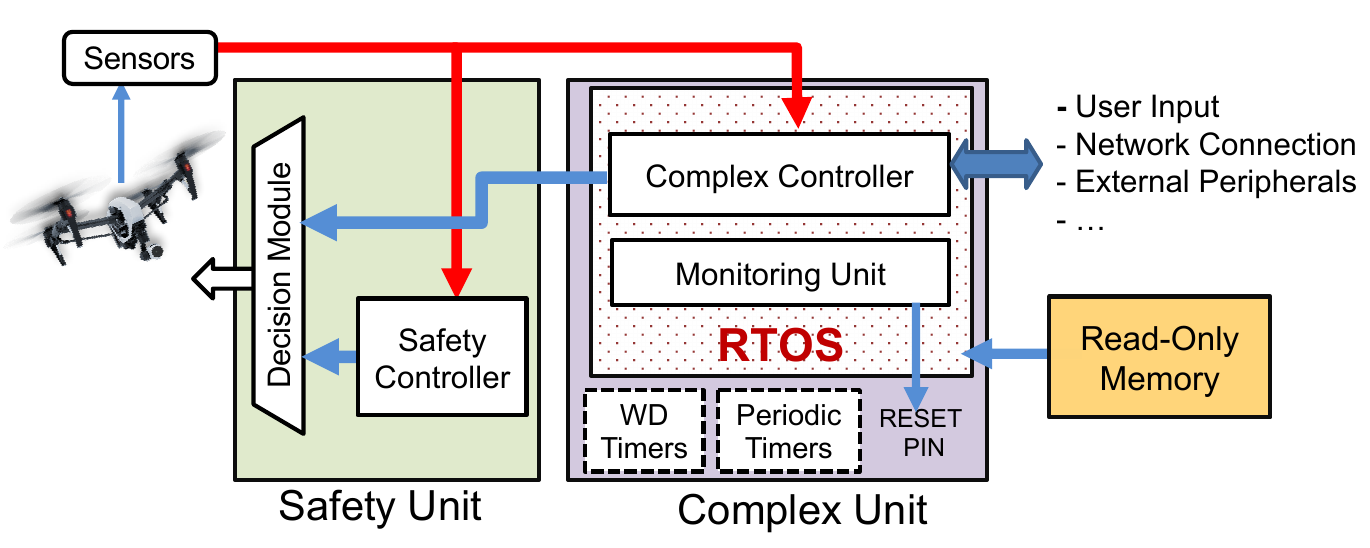}
\caption{The \textit{ReSecure} framework \cite{mhasan_resecure16}: safety unit is the bare-metal verified component, complex unit is not verified. The decision module switches between the controllers to provide overall system safety.}
\label{fig:resecure}
\end{figure}

Our primary focus here is to ensure the safety of the system despite the presence of malicious entity. The main idea is that, if we restart the system \textit{frequently enough}, it is less likely that the attacker will have time to re-enter the system and cause meaningful damage (such as data breaches and jeopardizing safety) to the system. After every restart, there will be a predictable down time (during the system reboot), some operational time (before system is compromised again) and some compromised time (until the compromise is detected or the periodic timer expires). 
The length of each one of the above intervals depends on the type and configuration of the platform, adversary models, 
complexity of the exploits, \etc~As a general rule, the effectiveness of the restarting mechanism increases: \ci as the time to re-launch the attacks increases, or \cii the time to detect attacks and trigger a restart decreases. We also evaluate the  expected lack of availability due to restarts and the expected damage from the attacks/exploits given a certain restart configuration. 

\begin{figure*}
	\centering
\includegraphics[width=0.8\textwidth]{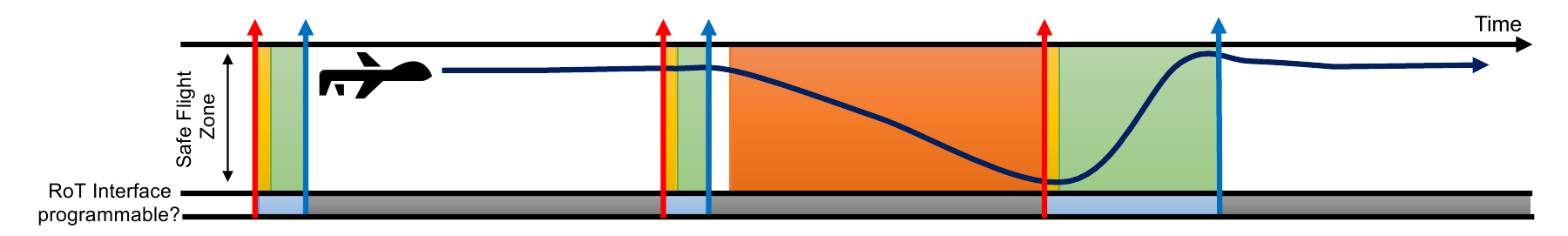}
	\caption{An example of a UAV system operating under the \textit{ReSecure} framework~{\cite{abdi2018guaranteed}}. The black line coming out from the UAV indicates the distance before it gets out of the safe flight zone. The red arrows annotate the triggering of the restarting points by the RoT. The blue arrows annotate the exit of the SEI (and that the next restart time is scheduled in RoT). We use different colors to illustrate the different phases of the system operation -- \ca \textit{white}: the main flight controller is in charge and system is not compromised; \cb \textit{yellow}: the system is restarting; \cc \textit{green}: SEI is active, the safety controller is running and the next restart time is being calculated; \cd \textit{orange}: the system is compromised and the adversary is in charge; \ce \textit{blue/gray}: the time spans when the (RoT) interface is available and unavailable, respectively. } 
	\label{fig:resecure_flow}
\end{figure*}

In later work~\cite{abdi2018guaranteed}, we introduced the secure execution interval (SEI) -- a period of time after each restart and before the untrusted applications begin to execute, when the execution environment is not yet contaminated and hence security is guaranteed.
During SEI, the system executes trusted code to determine the next restart time based on the current discrete state of the physical system.
When necessary, a safety controller can override the control of the system (during SEI) to guide the system back to a safe state.
In addition, we introduced a root of trust (RoT) -- an isolated hardware timer responsible for enforcing the restart process by issuing the restart signal at designated times (computed by the trusted code in SEI). RoT is designed to be programmable only once in each execution cycle and only during SEI.
Since it is inaccessible outside of SEI and works independently, the triggering of the restart process is not affected even when the system is compromised.
An example our framework operating in a UAV system is illustrated in Fig.~{\ref{fig:resecure_flow}}. The UAV operates normally within its safe flight zone and the safety controller does not need to be activated during SEI. Once the attacker compromises the system after the second restart (the orange area), the UAV flying towards its unsafe zone. Before the UAV reaches the unsafe zone, the hardware timer is up in RoT and triggers a restart. The safety controller (in SEI) takes over the control and brings the UAV back to the safe zone. Once the UAV returns to a predefined safe zone threshold, SEI ends and hands the control backs to the applications.

\subsection{Security without Architectural Modifications}

Despite the fact that architectural modification can improve the security posture of RT-IoT nodes, those approaches require an overall  redesign and may not be suitable for systems developed using COTS components. We now review the some of the approaches that we recently proposed to enhance security in RT-IoT without custom hardware support. 

\subsubsection{Dealing with Side-Channel Attacks}
\label{sec:secure-side-channel}
As introduced in Section~\ref{sec:cache-attack-demo}, we demonstrated that an attacker can carry out a cache-timing attack to indirectly estimate memory usage behavior.
It is due to the lack of isolation for shared resources across different tasks in most COTS-based RT-IoT systems.
The overlap between tasks happens when the system transitions from one task to another.
Therefore, capturing security constraints between tasks becomes essential for preventing side-channel attacks.

In previous work~\cite{sg1}, we proposed to integrate security in RT-IoT by introducing techniques to add constraints to tasks scheduled with fixed-priority real-time schedulers.
Based on user-defined security levels for each task, the scheduler \textit{flushes shared cache} when the system is transitioning from a high security task (\ie a task demanding higher confidentiality) to a low security task (\ie an insecure task potentially compromised). Let us consider the set of security
levels for tasks, $S$, that forms a {\em total order}. 
Hence, any two tasks ($\tau_i,
\tau_j$) may have one of  the following two relationships when
considering their security levels, $s_i,s_j \in S$: \ci $s_i \prec s_j$,
meaning that $\tau_i$ has higher security level than $\tau_j$ or \cii $s_j
\prec s_i$.

We proposed the idea of mitigating information leakage  
among tasks of varying security levels, by {\em transforming security requirements into constraints on scheduling algorithms}.
The approach of modifying or constraining scheduling algorithms is appealing because, 
\ca it is a software based approach and hence easier to deploy compared to hardware based approaches and
\cb it allows for reconciling the security requirements with real-time or schedulability requirements.
Consider a simple case with 
two periodic tasks, a high priority task $H$ and a low priority task 
$L$ scheduled by a fixed-priority scheduling policy. Assume that $s_H \prec s_L$; 
hence, information from H must not leak to L. These tasks must be scheduled on a 
single processor, $P$, so that both deadlines $(D_H, D_L)$ are satisfied. If $L$ (or any part thereof) executes immediately after (any part) 
or all of $H$, then $L$ could ``leak'' data from resources recently used by $H$. The main intuition is that a penalty must be paid for each shared resource in the system every time tasks switch between security levels. In this case, {\em the cache must be flushed before a new task is scheduled}. Hence, we proposed the use of an independent task, called the \textit{Flush Task} for this purpose.

In subsequent work \cite{sg2}, we relaxed many of the restrictions (\eg the requirement of total ordering of security levels) and proposed a {\em new, more general model} to capture security
constraints between tasks in a real-time system. This includes the analysis for the schedulability conditions with both preemptive and non-preemptive tasks.
We proposed a constraint named {\em noleak} to capture whether {\em
unintended information sharing between a pair of tasks must be forbidden}. Using this constraint we can prevent the information leakage via implicitly shared resources. For any two tasks $\tau_i$ and $\tau_j$: if $noleak(\tau_i, \tau_j) = \mathtt{True}$, then information leakage from $\tau_i$ to $\tau_j$ must be prevented; if $noleak(\tau_i, \tau_j) = \mathtt{False}$, no such constraints need to be enforced.
We showed that the system remains schedulable (\eg all the tasks can meet their deadline) under the proposed constraints without significant performance impact. 

%

\subsubsection{Schedule Randomization}
One way to protect a system from certain attacks (\eg the schedule-based side-channel attack mentioned in Section~\ref{sec:scheduleak}),
is to \emph{randomize the task schedule} to reduce the deterministic observability of periodic RT-IoT applications.
By randomizing the task schedules 
we can enforce non-determinism since every hyper-period\footnote{Hyper-period is the smallest interval of time after which the periodic patterns of all the tasks repeats itself -- typically defined as the least common multiple of the periods of the tasks.} will show different order (and timing) of execution for the tasks.
Unlike traditional systems, randomizing task schedules in RT-IoT is not straightforward since it leads to priority
inversions~\cite{Sha:1990:PIP} that, in turn, may cause missed deadlines --
hence, putting the safety of the system at risk.

Hence, we proposed \textit{TaskShuffler}~\cite{taskshuffler}, a randomization protocol for fixed-priority scheduling algorithm, to achieve such randomness in task schedule.
For instance, by picking a random task instead of the one with the highest-priority at each scheduling point, subject to the deadline constraints. The degree of randomness is flexible in \textit{TaskShuffler}. 
Based on the system's needs, \textit{TaskShuffler} implements the following randomization schemes:
\begin{itemize}
\item \textit{Randomization (Task Only)}: 
This is the most basic form of randomization in contrast to other schemes introduced below. 
We randomly pick a task to execute whenever a task arrives or finishes its job, \ie at the scheduling points. 
The effectiveness against the schedule-based side-channel attack is limited since the busy intervals in this scheme remains the same.

\item \textit{Randomization with Idle Time Scheduling}: 
In addition to the randomness provided in the basic scheme, we include the \textit{idle task} (\eg the dummy task executed by an RTOS when other real-time tasks are not running) at each scheduling point.
It eliminates the periodicity of busy intervals (from hyper-period's point of view).
This scheme makes it harder to produce effective results from the schedule-based side-channel attack.

\item \textit{Randomization with Idle Time Scheduling and Fine-grained Switching}: 
To push the randomization to an extreme, one could choose to randomize the schedule every tick.
That is, the scheduler will randomly pick a task to execute, subject to the deadline constraints, in every tick interrupt.
This way, we gain the most randomness for the schedule. Figure \ref{fig:taskshuffler} illustrates an instance of the randomized schedule for an simple taskset with three tasks.
However, it greatly increases 
the overheads and thus may not be applicable for all use cases.
\end{itemize}

\begin{figure}[h]
    \centering
   
    \begin{subfigure}[t]{0.49\textwidth}
        \centering
        \includegraphics[width=1.0\columnwidth]{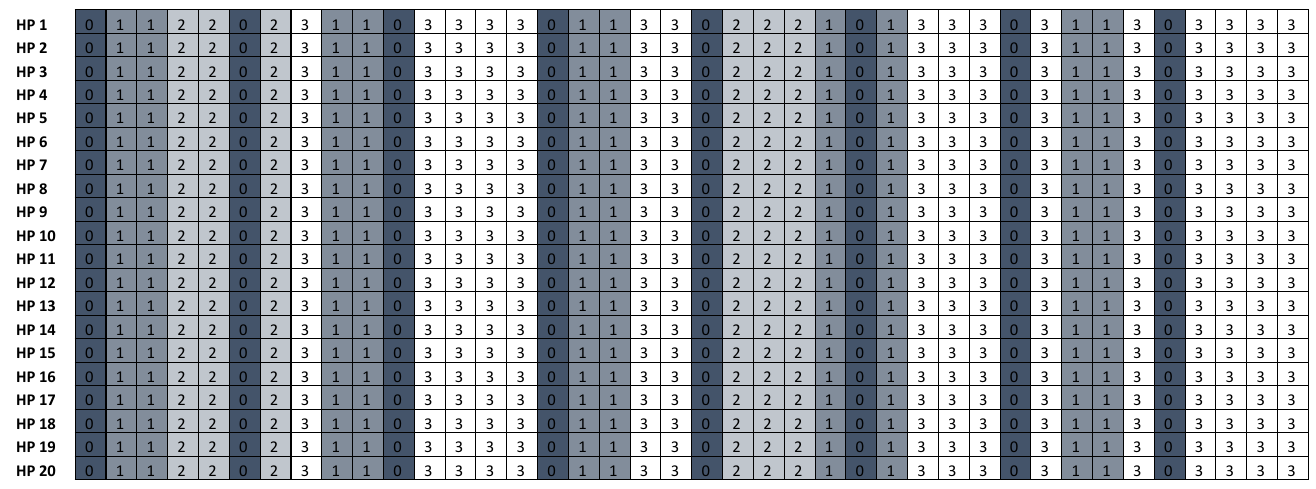}
        \caption{}
    \end{subfigure}%
    \hfill  
    \begin{subfigure}[t]{0.49\textwidth}
        \centering
        \includegraphics[width=1.0\columnwidth]{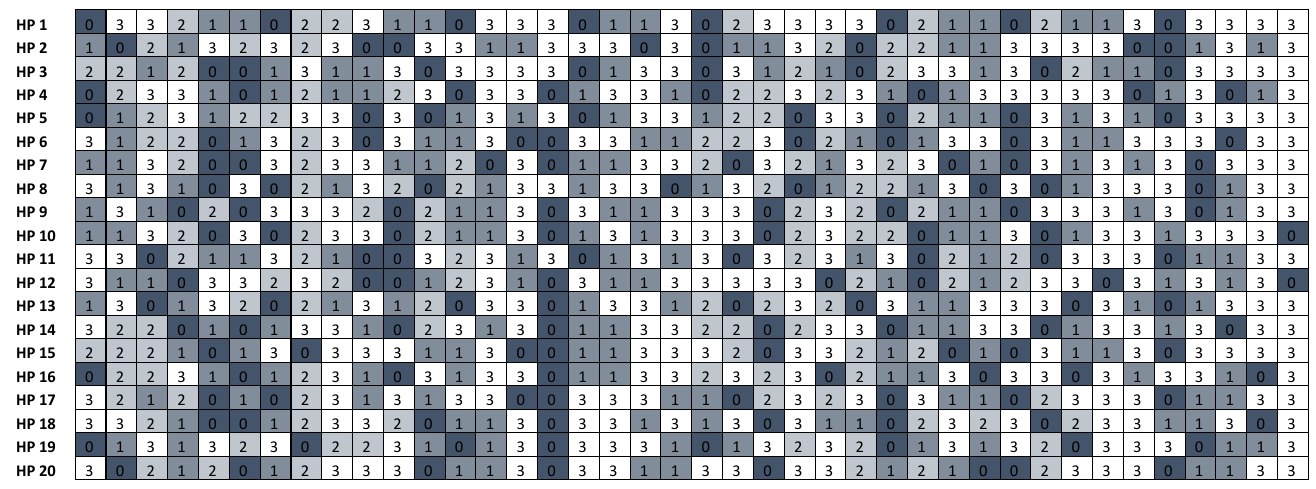}
        \caption{}
    \end{subfigure}
    \vspace{-\baselineskip}
    \caption{Examples of the schedule randomization protocol with three tasks: \ca vanilla fixed priority scheduling (\eg schedules without randomization); \cb  \textit{TaskShuffler} (fine-grained scheduling with randomizing idle times). Blocks numbered 0 to 2 are the execution of periodic tasks while blocks numbered 3 indicate idle time (\ie the idle task). The following taskset parameters are used in the illustration: $\tau_0(5, 1, 5)$, $\tau_1(8,2,8)$, $\tau_2(20,3,20)$ where each task $\tau_i(C_i, T_i, D_i),~0 \leq i \leq 2$ is characterized by WCET ($C_i$), period ($T_i$) and deadline ($D_i$). Each row represents a hyperperiod and the figure shows the schedule of 20 hyperperiod.  For vanilla scheduling, task schedules are repeating each hyperperiod. In contrast, \textit{TaskShuffler} scrambles the schedule across hyperperiod and thus make it harder to predict a particular task execution instance. }
    \label{fig:taskshuffler}
\end{figure}

IoT systems with real-time properties are predictable by design. This very
determinism can become a vulnerability in the hands of smart adversaries and it becomes easier to carry out adversarial actions such as side-channel attacks \cite{cy_side_channel,side_channel_key},
DoS (making critical resources unavailable at important times) or
even the recently developed timing-inference attacks \cite{cy_side_channel}.~\textit{TaskShuffler} can reduce the determinism that is 
visible to external entities while still meeting real-time guarantees. With such randomization, even if an observer is able to capture the exact schedule for a (limited) period of time (for instance, for a few hyperperiods), \textit{TaskShuffler} will schedule tasks in a way that succeeding hyperperiod will show different orders (and timing) of execution for the tasks.

\subsubsection{Integrating Security for Legacy RT-IoT}  \label{subsec:sec_int}


As we have described in Section \ref{sec:scheduleak}, an adversary can
extract important information while still remaining undetected 
and it is essential to have a layered
defense and integrated resilience against such attacks into the design of RT-IoT.
However, any security mechanisms have to \textit{co-exist} with real-time tasks in
	the system and have to operate without impacting the timing and safety
	constraints of the control logic. Besides, the embedded nature of these systems limits the availability of computational power (\eg memory or processor) required for resource-extensive monitoring mechanisms.
    This creates an apparent tension between
security requirements (\eg having enough cycles for effective monitoring and 
detection) and the timing and safety requirements. For example, a critical parameter is to determine how often and
how long should a monitoring and intrusion detection task execution to be effective but not
interfere with real-time control or other safety-critical tasks. While this
tension could potentially be addressed for newer systems at design time, this
is especially challenging for retrofitting \textit{legacy} systems where the control
tasks are already in place and perhaps \textit{cannot be modified}. Any hardware and/or software-level modifications to those legacy system parameters is costly since it will go through several verification and validation steps and may increase system downtime \cite{iot_fog_overview}. 
Most of the security solutions for RT-IoT proposed in literature either require custom hardware \cite{mohan_s3a, securecore,securecore_memory,securecore_syscal,onchip_fardin, mhasan_resecure16,slack_cornell,abdi2018guaranteed}, modification of the existing schedulers \cite{xie2007improving, lin2009static}, extra instrumentations \cite{slack_cornell} or may need to change the tasks parameters (\eg execution order and/or run-time) \cite{sg1, sg2,taskshuffler} and therefore \textit{not} suitable for legacy systems. Integrating monitoring and
detection tasks for RT-IoT \textit{without} custom hardware support is an open problem.

Given the tension between security and timing requirements, while integrating
security mechanisms into a practical system, 
finding the \textit{frequency of execution} of the monitoring tasks is an important design parameter that trades security requirements with timing constraints. 
If the interval
	between consecutive monitoring events is too large, the adversary may
	harm the system (and remain undetected) between two invocations of the
	security task. In contrast, if the security tasks are executed
	very frequently then it may impact the schedulability of the real-time
	tasks. 

In preliminary work \cite{mhasan_rtss16} we address 
the problem of determining the frequency of execution (\eg periods or inter-monitoring interval) of the security tasks. Our approach to integrate security without perturbing real-time scheduling order
is to execute security tasks at a \textit{lower priority} tasks than real-time tasks. We refer this scheme as \textit{opportunistic execution} since the security tasks are only allowed to execute opportunistically only during \textit{slack times} when no
other real-time tasks are running.

We propose to measure the security of the system by means of the \textit{achievable
periodic monitoring}. Let $T_i$ be the period of the security task 
that needs to be determined. Our goal here is to minimize the
perturbation between the achievable (\ie unknown) period $T_i$ and the desired (\eg designer provided) period
$T_i^{des}$. 
We formulate a constraint optimization problem and develop a polynomial-time solution that allows us to execute security
tasks with a frequency closer to the desired values while respecting the
temporal constraints of the other real-time tasks. 


\begin{figure}[t]
\centering
\includegraphics[scale=0.27]{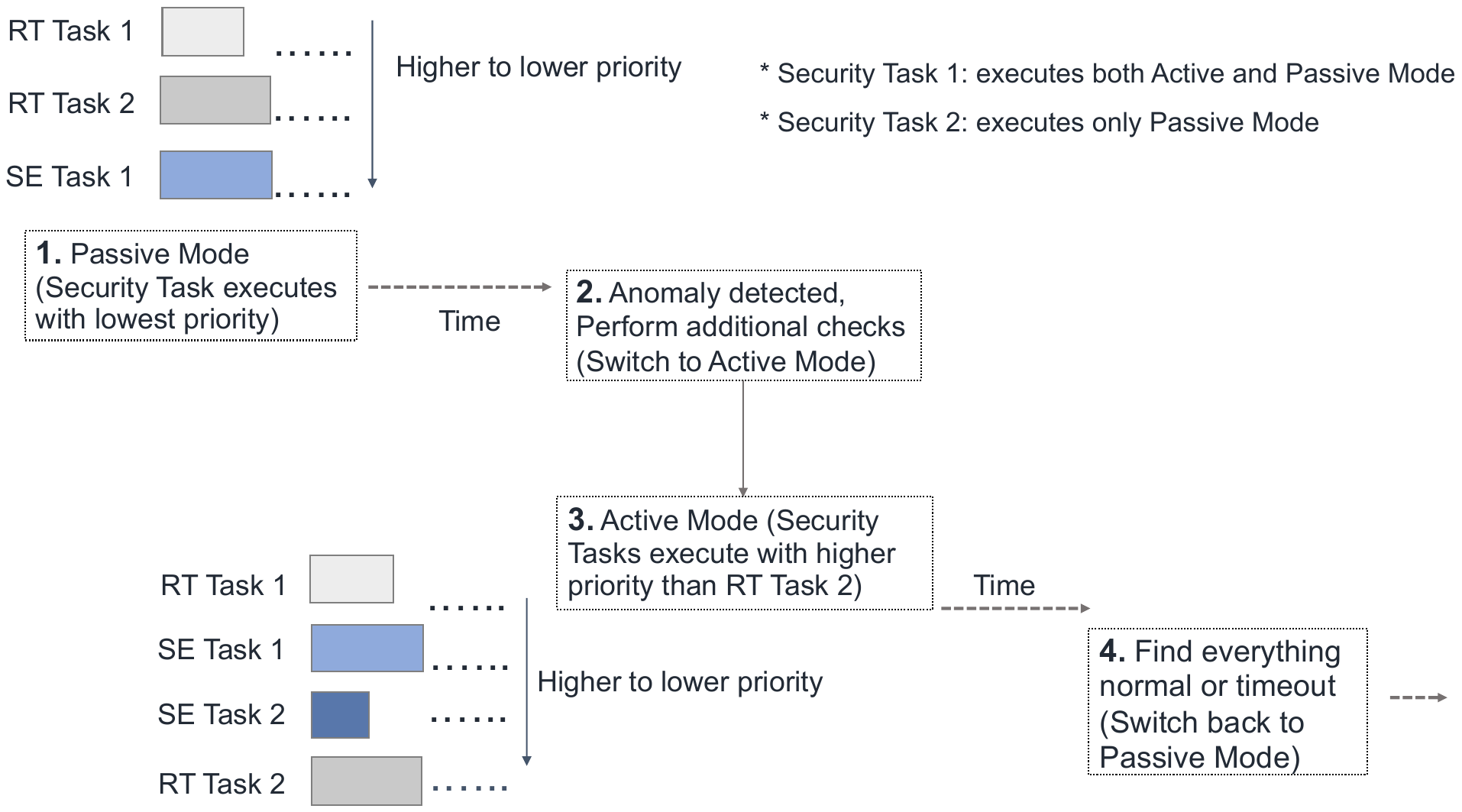}
\caption{Flow of operations in \textit{Contego} depicting different modes for the security tasks.}
\label{fig:contego}
\end{figure}

If the security tasks always execute with lowest priority, they suffer more interference (\ie preemption from high-priority real-time tasks) and the consequent longer detection time (due to poor response time) will make the security mechanisms less effective. In order to provide \textit{better responsiveness} and increase the effectiveness of monitoring and detection mechanisms, we then proposed a multi-mode model \cite{mhasan_ecrts17}. This framework (called  
\textit{Contego}) allows the security policies/tasks to execute in different modes in different \textit{modes} (\ie passive monitoring with lowest priority as well as exhaustive checking with higher priority). By using this approach (see Fig. \ref{fig:contego}), for instance, security routines can execute opportunistically when the system is deemed to be clean (\ie not compromised). However if any anomaly or unusual behavior is suspected, the security policy may switch to a \textit{fine-grained checking mode} and execute with higher priority. The security routines may go back to normal mode if: \textit{i}) no anomalous activity is found; or \textit{ii}) the intrusion is detected and malicious entities are removed.

The aforementioned works however are developed for single core systems only -- integrating security mechanisms for legacy multicore platforms (where designers have less flexibility for changing system architecture/parameter) is also a challenging problem. In recent work \cite{mhasan_date18} we developed a 
scheme for multicore RT-IoT and find a suitable assignment of security tasks that ensures they can execute with a frequency close to what a designer expects. We considered a multicore platform comprised of $M$ identical
cores. One fundamental problem while integrating security mechanisms in multicore platforms is to determine \textit{which security tasks will be assigned to which core and executed when}.
Although security tasks can execute in any of the $M$ available cores and any period $T_i^{des} \leq T_i \leq T_i^{max}$ is acceptable, the actual task-to-core assignment and the periods of the security tasks are not known apriori. The goal of this scheme therefore is to jointly find the \textit{core-to-task assignment and suitable periods} for the security tasks. However, finding such an assignment is NP-hard due to combinatorial nature of the problem. Therefore we developed a
\textit{near-optimal low-complexity solution} (called \textit{HYDRA}) that jointly finds the security tasks' period and core assignments. From our experiment we found that
on average \textit{HYDRA} (that distributes security tasks across all available cores) can provide 27.23\% faster intrusion detection rate (on a quad core system) compared to the case when the security tasks are allocated a dedicated core for
while the real-time tasks are assigned to the remaining
cores.

\section{Discussion and Research Opportunities}





\subsection{Securing Legacy RT-IoT Systems}
\label{sec:discussion}

 Since most RT-IoT nodes are resource-constrained embedded devices, resource-intensive processing
and complex protocols (\eg heavy cryptographic operations) for securing those systems is unrealistic and may threaten the safety of such systems -- for instance a safety-critical task may miss deadline in order to run computation-heavy security tasks.
 In addition to execution frequency, another important consideration is to determine how quickly can intrusions be detected. Thus \textit{responsiveness vs.
	schedulability} of critical tasks is another important trade-off. This in itself is a research
challenge that needs to be investigated.

So far we have assumed that we are given a set of security tasks and that each
security task has a desired frequency of execution for better security
coverage. Security tasks so far have been treated as being independent and
preemptible. But in practice, as previously discussed some security monitoring
may need atomicity or non-preemptive execution. Further, \textit{security tasks may
have dependencies where one task depends on the output from one or more other
tasks}. For example, an anomaly detection task may depend on the outputs of
multiple scanning tasks. Or, the scheduling framework may need to follow
certain \textit{precedence constraints} for security tasks. For example, in
order to ensure integrity of monitoring security, the security application's
own binary may need to be examined first before checking the system binary
files. In such cases we can not independently execute the security task and we need to consider the problem of integrating security
tasks with dependencies between them. One approach could be use a directed
acyclic graph (DAG) to capture the dependencies and constraints among security
tasks. In this case, tightness of achievable
periodic monitoring described in Section~\ref{subsec:sec_int} may no
longer be a reasonable metric. Constraints to ensure that the entire DAG is
executed often enough should be included and the optimization problem
reformulated and evaluated with different metrics.

\subsection{Security for Multicore based RT-IoT Platforms}


Most of the work \cite{chen2015schedule, sg1,sg2,mhasan_rtss16, mhasan_ecrts17} presented so far has been in the context of single core processors
-- they are the most common types of processors being used in RT-IoT
systems. However, as mentioned earlier, due to increasing computational demands, multi-core
processors are becoming increasingly relevant to real-time systems \cite{iot_wind_river_multicore, multicore_survey}.
With the increased number of cores, more computation can be packed into a
single chip -- thus reducing power and weight requirements -- both of which
might be relevant to many RT-IoT systems. However multicore processors can {\em increase attack vectors},
especially for side-channel attacks.  First, two or more tasks are
running together and (most likely) sharing low-level resources (\eg last level
caches). Hence, a task running on one core can snoop on the other -- and not
just when tasks follow each other. In fact, it has been shown that leakage can occur
with a much higher bandwidth in the case of shared resources in multi-core
processors \cite{gesurvey}. Second, when tasks execute together, a malicious task can
increase the ``interference'' faced by a critical task -- for instance, the
malicious task can flood the cache/bus with memory references just when an
important task (say, one that computes the control loop) is running.  This
could cause the critical task to get delayed and {\em even miss its deadline}. To prevent such problems, designers of the systems need to enforce constraints that protected tasks
do not execute simultaneously with unprotected ones on the multi-core chip.

The problem of integrating security tasks into legacy RT-IoT systems is also interesting in the multicore context -- perhaps the security tasks can always
	be running (say on one of the dedicated cores) instead of running opportunistically as is the case for single core systems. Also it may be possible to
	to take up more cores and execute \textit{fine-grained sanity checks} (\eg a complete system-wide scan) as it detects malicious activity. Analyzing the impact of integrating security tasks in a multicore legacy RT-IoT is an open problem worth investigating.



\subsection{Secure Communication with Timing Constraints}

With the rise of RT-IoT, the edge devices are more frequently exchanging control messages and data often with unreliable mediums like the Internet. Therefore, in addition to the host-based approaches \cite{mohan_s3a,securecore,securecore_memory,securecore_syscal,onchip_fardin,taskshuffler,sg1,sg2} described earlier, there is a requirement for securing communication channels to ensure authenticity and integrity of control messages. While some of our previous work \cite{mhasan_resecure16,mhasan_rtss16} can also be used to deal with network-level attacks, designing a unified framework to protect edge devices as well as communication messages (given the stringent end-to-end delay requirements for high-critical traffics) is 
still an open problem. 

Most safety-critical RT-IoT systems often have separate networks (hardware and
software) for each of the different types of flows for safety (and security) reasons. This leads to significant overheads (equipment, management, weight, \etc) and
also potential for errors/faults and even increased attack surface and vectors. Network-level nondeterminism, \ie unpredictability in sensor reading, packet delivery/forwarding/processing further complicates the management of RT-IoT systems.
Existing protocols, \eg avionics full-duplex switched Ethernet (AFDX) \cite{afdx_evolution}, controller
area network (CAN) \cite{can_overview}, \etc~that are in use in many of real-time domains are either
proprietary, complex, expensive, require custom hardware or they are also exposed to known vulnerabilities \cite{can_security}. 

Given the limitations of existing protocols,  
leveraging the benefits of \textit{software-defined networking} (SDN) can also be effective for RT-IoT systems. The advantage of using SDN is that it is compatible with COTS components (and thus suitable for legacy RT-IoT systems) and provides a centralized mechanism for
developing and managing the system. The global view is
useful to ensure QoS (\eg bandwidth and delay) and enforce security mechanisms (such as remote attestations, secure key/message exchange, remote monitoring, \etc). While SDNs provide a global view of the network and high-level management
capabilities (including resource allocation), current standards used in traditional SDN (\eg OpenFlow \cite{mckeown2008openflow}) do not consider inherent timing and safety-critical nature of the RT-IoT systems. In recent
work~\cite{sdn_qos_rtss17} we tried to address this problem
through static flow allocation and routing -- we
used static path allocation and over-provisioning hardware resources (\eg dedicating one queue per real-time flow) for
meeting the end-to-end delay requirements and providing isolation. This limited
the number of flows that could be admitted and resulted in underutilized network
resources. Retrofitting the capabilities of SDN in the RT-IoT domain requires further research. We also need mechanisms to \textit{prioritize between flows}
(say between the critical real-time flows or even across real-time and non real-time flows) and also
 schemes for \textit{multiplexing flows} 
on the same queues in the SDN switches (to improve the efficiency of the network) while still meeting the real-time constraints.


\section{Related Work}

There exists work that has investigated security in real-time systems~\cite{xie2007improving,lin2009static,SonCT98}. 
Many researchers have studied this research area from different aspects.
Information leakage via side channels has been discussed in many works.
Kadloor \etal \cite{kadloor2013} and Gong \etal \cite{GongK14} introduced analysis and methodology for quantifying side-channel leakage.
Kelsey \etal \cite{kelsey1998side}, Osvik \etal \cite{osvik2006cache} and Page \etal \cite{page2002theoretical} demonstrated the usability of cache-based side-channels.
Son \etal \cite{embeddedsecurity:son2006} and V\"{o}lp \etal \cite{embeddedsecurity:volp2008} examined the exploitation of timing channels in real-time scheduling.
 Bao \etal~\cite{thermal_profile_ccs} introduce a scheduling algorithm for soft real-time systems (where some tasks can miss deadlines) and provide a trade-off between thermal side-channel information
leakage and the number of deadline misses. Their exists other work \cite{side_channel_key} that studies the robustness of AES secret keys against
differential power analysis (DPA) \cite{differential_power_intro} attacks.

While the work above focuses on exploring vulnerabilities, there exist work that aims to provide security to real-time systems.
Ghassami \etal~\cite{ghassami2015capacity} and V\"{o}lp~\etal \cite{embeddedsecurity:volp2013} proposed techniques to address leakage via shared resources. 
An online job randomization scheme~\cite{kruger2018vulnerability} is proposed by Kr\"{u}ger \etal
for time-triggered real-time systems.
Xie \etal~\cite{xie2007improving} and Lin \etal~\cite{lin2009static} presented security in real-time systems by encrypting communication messages. Similar to the hardware-assisted security mechanisms like ours (\eg \textit{S3A}, \textit{SecureCore}, \textit{ReSecure}, \etc) researchers also propose architectural frameworks \cite{slack_cornell} that dynamically utilizes slack times (\eg the time instance when no other real-time tasks is executing) for run-time monitoring. There exists recent work \cite{lesi2017network, lesi2017security} where authors proposed schemes to secure systems from man-in-the-middle attacks, where an attacker can compromise communication between system sensors and controllers.

Some recent work has raised security awareness in IoT applications \cite{ida2016survey, ngu2016iot, mohsin2016iotsat, kraijak2015survey, wurm2016security}.
Some researchers aim to add security properties to IoT.
Pacheco \etal \cite{pacheco2016iot} introduced a security framework that offers security solutions with smart infrastructures.  
Kuusij{\"a}rvi \etal~\cite{kuusijarvi2016mitigating} proposed to mitigate IoT security threats with using trusted networks.
Those work primarily focuses on generic IoT applications, and do not consider the additional real-time constraints required for RT-IoT systems.

\section{Conclusion}

The sophistication of recent attacks on RT-IoT requires rethinking of security solutions for such systems.
The goal of this paper is to raise the awareness of real-time security and bridge missing gaps 
in the current IoT context -- securing the IoT systems with real-time constraints.
The techniques and methodology presented here vary from different perspectives -- from hardware-assisted security to scheduler-level as well as those for legacy systems. The designers of the systems and research community will now be able to integrate and develop upon these frameworks required to secure safety-critical RT-IoT systems.
We believe that the real-time and IoT worlds are closely connected and will become inseparable in the near future. 

\section*{Acknowledgement}

The multiple security frameworks presented in this paper is the result of a team effort. The authors would like to acknowledge the contributions from our team members and collaborators: Man-Ki Yoon, Fardin Abdi Taghi Abad, Rodolfo Pellizzoni, Rakesh B Bobba, Lui Sha and Marco Caccamo. This work is supported in part by grants from the National Science Foundation (NSF CNS 14-23334, NSF CPS 1544901 and NSF SaTC 1718952), ONR (N00014-13-1-0707) and the Dept. of Energy.  Any opinions, findings, and conclusions or recommendations expressed here are those of the authors and do not necessarily reflect the views of sponsors. 




\bibliographystyle{IEEEtran}

\bibliography{references_short_mhasan,references_cy_short}


\end{document}